\documentstyle[epsf]{aipproc}
\begin{document}
\title{
Stripe Disordering Transition}

\author{R.S. Markiewicz$^{1,2}$ and M.T. Vaughn$^1$} 

\address{Physics Department (1) and Barnett Institute (2), 
Northeastern U.,
Boston MA 02115}
\maketitle
\begin{abstract}
We have recently begun Monte Carlo simulations of the dynamics of stripe phases
in the cuprates.  A simple model of spinodal decomposition of the holes allows
us to incorporate Coulomb repulsion and coherency strains.  We find evidence
for a possible stripe disordering transition, at a temperature below the 
pseudogap onset.  Experimental searches for such a transition can provide 
constraints for models of stripe formation.

\end{abstract}

The relationship between stripe phases and the pseudogap in underdoped cuprates
is not well understood.  In our model\cite{Pstr,MKK,Mia} the pseudogap is
primary.  It represents an instability of the hole Fermi liquid driven by
Van Hove nesting\cite{RiSc}.  However, there is a competition of instabilities,
with an antiferromagnet (or flux phase\cite{Affl,Laugh,WeL}) at half filling
and a charge-density wave (CDW) at the bare Van Hove singularity (VHS) near 
optimal doping.  This competition leads to a classical phase separation of the
holes -- two minima in the free energy\cite{RM3,Pstr}.  This is restricted to
a nanoscopic scale by long-range Coulomb effects, leading to phases similar to
the experimentally observed stripe phases\cite{Tran}.

For such a nanoscale phase separation, the correct dispersion and pseudogap 
must be found by appropriate averaging over the heterogeneous, usually 
fluctuating stripes.  Fortunately, 
tunneling and photoemission are sensitive mainly to the pseudogaps, and hence
can be described by a simple {\it Ansatz} of the stripe phase\cite{MKK,Mia}.
For other purposes, a more detailed picture of the stripes is needed.  

\begin{figure}
\leavevmode
   \epsfxsize=0.46\textwidth\epsfbox{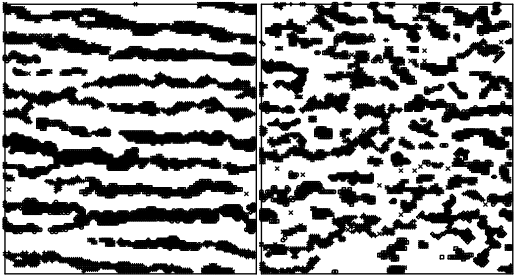}
\vskip0.5cm 
\caption{Monte Carlo calculated striped phases, with $x=0.06$, at two
temperatures: a low T (left), and near the melting point of the stripes 
(right). Orthorhombic axes, $\epsilon =180$.}
\label{fig:11}
\end{figure}

\begin{figure}
\leavevmode
   \epsfxsize=0.33\textwidth\epsfbox{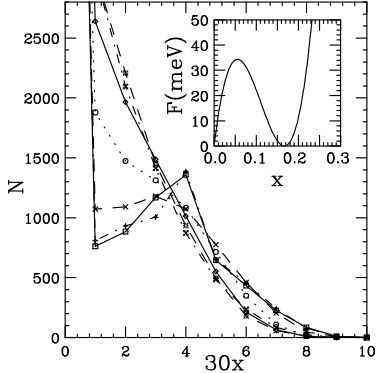}
\vskip0.5cm 
\caption{Striped phase melting transition at x=0.04, $\epsilon =23$: hole 
distribution function at temperatures (from bottom to top, at $x=2/30$) $k_BT$ 
= 10, 0.1, 30, 60, 100, 400, 200 meV.  Inset: Assumed free energy vs. doping.}
\label{fig:13}
\end{figure}

As a first step, we have begun Monte Carlo calculations of a classical picture 
of this restricted phase separation. Using the derived form of the free energy
vs doping, we calculate the dynamic spinodal 
decomposition of the holes in the presence of Coulomb interactions.  We find
that there can be a stripe disordering transition, Fig. \ref{fig:11}, at a
temperature below the pseudogap onset.  The disordering temperature is 
proportional to the free energy barrier between the two end-phases, inset,
Fig. \ref{fig:13}.

Technical details of the calculation are as follows: we work with a generic
form of the free energy, $F=F_0x(x-x_c)^2$, which approximates the calculated
free energy of Ref. \cite{Pstr}.  The calculations are done on 128$\times$128
lattices, with periodic boundary conditions.  The critical doping $x_c$ is 
taken as 1/6, which necessitates a non-Markovian algorithm -- a particular 
lattice site must retain memory of the average hole occupation over several 
cycles. We typically choose 30 cycles, which means that a single hole must 
spread out over 6 lattice sites -- close to the size of a magnetic 
polaron\cite{Auer}.  The algorithm chosen is able to find the correct ground 
states in the low doping limit (which can be found analytically).  The stripes 
are not topological, and the stripe-like domains are produced by coherency 
strains\cite{FraPe}.  In the absence of such strains, the domains would be
irregular shaped, approximately equiaxed, as found by Veillette, et 
al.\cite{VBBK}  The coherency strains produce a mixture of stripes along both
$x$ and $y$ axes; to get single-axis stripes, as in the figure, it is assumed
that there are local martensitic domains.  


The phase separation can be most clearly seen in a plot of the distribution of
site occupancies by holes, Fig. \ref{fig:13}.  At low temperatures, this is a
two-peaked structure, with one peak (off scale in the figure) at zero doping,
and the other near $x_c$ (it is actually at a doping below $x_c$, due to 
charging effects).  As the temperature increases, the two-peak structure is
gradually smeared out, and at high temperatures there is only a monotonic
distribution.  This finite system has a crossover rather than a sharp 
transition.  For the parameters chosen, the transition is centered near $k_BT_m
\sim 30meV$, which is approximately the barrier height of the free energy 
(inset).  This result is not very sensitive to the value of dielectric constant,
$\epsilon$.

Thus, as the underdoped cuprate cools from high temperatures, there can be a 
series of phase transitions.  At high temperatures, there will be the
pseudogap onset.  In our simplified mean field {\it Ansatz}\cite{Mia}, this 
appears as a long-range ordered CDW phase, but the inclusion of two-dimensional 
fluctuations\cite{KaSch,RM5} leads to appropriate pseudogap behavior.  The
stripe phase ordering temperature found here could in principle fall at a lower
temperature.  The stripes in our simulations continue to fluctuate, and the
long-range stripe order phase seen by Tranquada\cite{Tran} may be yet another
transition.  The two-branched transition to a stripe phase bears some 
resemblance to the phase diagram of Emery, Kivelson, and Zakhar\cite{EKZ}, but
is in fact different.  Their upper transition ($T_1^*$) corresponds to the onset
of stripe order, their lower ($T_2^*$) to the onset of a spin gap on the hole
doped stripes.

There is not much
experimental evidence for the onset of short-range stripe order, although
phase separation in La$_2$CuO$_{4+\delta}$ starts near 400K\cite{Rad}, much
lower than the pseudogap onset temperature, $\sim 800K$\cite{BatT}.  In most
materials, the incommensurate magnetic modulations near $(\pi ,\pi )$ broaden
out and disappear near the pseudogap $T^*$, which is a lower temperature
($\sim\le$150K for the compositions studied)\cite{Moo}.  The best place to look
would be in the extremely underdoped regime, where $T^*$ is highest.

While the above calculations reproduce the general properties of the stripes, 
there are a number of features which are not well reproduced.  First, for the
elastic constants of LSCO\cite{Mig}, the stripes lie along the orthorhombic axes
-- i.e., they are diagonal stripes.  Further, for the parameters assumed, the
charged stripes tend to grow wider with increased doping, maintaining a
constant interstripe spacing, whereas experiment\cite{Yam} suggests that the
stripe shape stays constant, but the stripes move closer, as doping increases,
at least for $x\le 0.12$.  This suggests that some important feature has been
omitted from the model, most probably the topological nature of the stripes as
magnetic antiphase boundaries.

MTV's work was supported by DOE Grant DE-FG02-85ER40233.  Publication 758 of 
the Barnett Institute.

\end{document}